\documentclass{emulateapj}

\slugcomment{In press for ApJ Letters, December 20, 2007}
\shorttitle{HD 15745 debris disk}
\shortauthors{Kalas et al.}

\begin{document}

\title{Discovery of an extended debris disk around the F2V star HD 15745}

\author{Paul Kalas\altaffilmark{1,2}, Gaspard Duchene \altaffilmark{1,3}, 
Michael P. Fitzgerald\altaffilmark{2,4,5}, James R. Graham\altaffilmark{1,2}}
\affil{}

\altaffiltext{1}{Astronomy Department and Radio Astronomy Laboratory, 
601 Campbell Hall, Berkeley, CA 94720}
\altaffiltext{2}{National Science Foundation Center for Adaptive Optics, University of California, Santa Cruz, CA 95064}
\altaffiltext{3}{Laboratoire
d'Astrophysique de Grenoble, Universit\'e Joseph Fourier, BP 53, 38041
Grenoble cedex 9, FRANCE}
\altaffiltext{4}{Lawrence Livermore National Laboratory, 
7000 East Ave., Livermore, CA 94551}
\altaffiltext{5}{Michelson Fellow}

\begin{abstract}
Using the Advanced Camera for Surveys aboard the {\it Hubble Space Telescope},
we have discovered dust-scattered light from the debris disk surrounding
the F2V star HD 15745.  The circumstellar disk is detected between 2.0$\arcsec$ and
7.5$\arcsec$ radius, corresponding to 128$-$480 AU radius.  
The circumstellar disk morphology is asymmetric about the star, resembling a
fan, and consistent with forward scattering grains in an optically thin
disk with an inclination of $\sim$67$\degr$ to our line of sight.   The spectral
energy distribution and scattered light morphology can be approximated
with a model disk composed of silicate grains between 60 and 450 AU radius, 
with a total dust mass of $10^{-7}~M_\odot$ (0.03 M$_\earth$) representing
a narrow grain size distribution (1$-$10 $\mu$m).  
Galactic space motions are similar to the Castor Moving
Group with an age of $\sim$10$^{8}$ yr, although future work is required to determine
the age of HD 15745 using other indicators. 
\end{abstract}

\section{Introduction}
The collisional debris from solid bodies orbiting main sequence stars results in circumstellar dust
disks that produce excess emission at thermal infrared wavelengths \citep{bap93}.  
Excess thermal emission may also arise from a star 
heating interstellar dust \citep[e.g.,][]{kalas02} and therefore obtaining
images of dust-scattered light is a key method for constraining the origin of dust.  Furthermore,
direct images of dust-scattered light may be used to obtain the color and polarization properties
of the grains, which depend on their size distribution, composition and structure \citep[e.g.][]{graham07}.  
The overall geometry of a debris disk may also reveal sources of dynamical perturbations, such as those
from stellar and sub-stellar companions \citep[e.g.][]{roques94, liou99, moro02}.   

Here we show the first scattered-light images of a debris disk surrounding HD 15745, 
an F2V star at 64 pc.  \citet{silverstone00} first reported HD 15745 as
a debris disk candidate star based on excess thermal emission 
$L_{IR}/L_{\star} = 1.8\times10^{-3}$ using $IRAS$ data, and 
$L_{IR}/L_{\star} = 1.3\times10^{-3}$ using $ISO$ data.   \citet{zuck04} modeled
the spectral energy distribution (SED) with a single temperature dust 
belt at 27 AU radius with $L_{IR}/L_{\star} = 1.2 \times 10^{-3}$.

\section{Observations \& Data Analysis}
We observed HD 15745 on 2006 July 17 with the {\it Hubble Space Telescope (HST)}
Advanced Camera  for Surveys (ACS) High Resolution Channel (HRC) coronagraph.  
The star was acquired behind the 1.8$\arcsec$ diameter occulting spot and three 700 second
integrations with the F606W ($\lambda_c$= 591 nm, $\Delta\lambda$ = 234 nm)
broadband filter were obtained at a fixed position angle.  
The three frames were median combined for cosmic ray rejection.  
The point spread function (PSF) was then subtracted using images from
the HST archive of five F stars obtained with the ACS coronagraph
at the same wavelength and comparable integration times.
All five versions of a PSF-subtracted image of HD 15745 show the
fan-shaped circumstellar nebulosity.   The five versions
were then median combined to produce a final image that
was then corrected for geometric distortion.
We also detected HD 15745 from Gemini North with the MICHELLE
camera on 2006 September 20 using the N$^\prime$ filter.  
We measure a flux of N$^\prime$= 74.7$\pm$2.9 mJy within a 1.4$\arcsec$ radius
aperture, with an aperture correction based on near-contemporaneous 
images of the photometric standard star (HD 20893). 

\section{Results}
The PSF-subtracted HST data reveal a fan-shaped nebulosity surrounding HD 15745 in the
approximate region between P.A.=190$\degr$ and P.A.=10$\degr$ (Fig. 1).  
We interpret the nebulosity as
an  azimuthally symmetric circumstellar disk inclined 
to our line of sight  and composed of dust that has an asymmetric scattering 
phase function that produces a  fan-like appearance \citep{kalas96}.
Another possibility $-$ asymmetric grain erosion and blowback by encounters with
the interstellar medium \citep{arty97}$-$ is less likely given that the southeastward proper
motion ($\mu_\alpha$=46 mas/yr, $\mu_\delta$=-47 mas/yr) and location of the star well
within the local interstellar bubble \citep{kalas02}, although at least one exception has been
found for the latter point \citep{gaspar07}.  

In principle the detection 
limit for the inner radius of the nebulosity is set by the occulting spot radius (0.9$\arcsec$), but
the presence of quasi-static speckles produce significant PSF subtraction
artifacts, limiting the inner detection radius to $\sim$2$\arcsec$ (128 AU) radius.  The 
outer radius of the disk is a sensitivity-limited value.   In data that have been binned 8$\times$8
pixels (0.2$\arcsec\times0.2\arcsec$), the disk is detected as far as $\sim$7.5$\arcsec$ (480 AU) radius.  
The 23.0 mag arcsec$^{-2}$ surface brightness isophote (Fig. 2) shows a relatively symmetric fan-like
morphology.  The radial surface brightness profiles (Fig. 3) are monotonic and can be
approximated by power laws with an index of $-3.3\pm0.1$.

\section{Disk models}

We model the SED and scattered light appearance of the HD 15745 disk
using a Monte Carlo radiation transfer code (MCFOST; Pinte et
al. 2006) in which a star radiates isotropically in space and illuminates
an azimuthally-symmetric parameterized disk. In a first step, we
determine the stellar parameters by fitting the SED up to 11\,$\mu$m (Fig. 4). 
Using NextGen models \citep{baraffe98} for the
stellar photosphere, we find that the best fit model has
$T_{eff}=6800$\,K and $R_\star=1.31R_\odot$.

We then proceed to fit the entire SED of HD 15745 to determine the
disk properties. In order to consider as few free parameters as
possible, we adopt a $\rho(r,z) \propto r^\alpha \exp(-z^2/2\sigma_z^2)$
number density distribution, with $\alpha=-1$ and a constant
$\sigma_z=2$\,AU, that extends from $R_{in}$ to $R_{out}=450$\,AU and
sums up to a total disk mass $M_d$. The choice of $\sigma_z=2$\,AU
is consistent with disk heights inferred from the edge-on debris
disks $\beta$ Pic and AU Mic \citep{kalas95, krist05}, and does
not significantly influence our results for values $\sigma_z\geq1$\,AU
(which is required to keep the disk optically thin to starlight).
The $r^{-1}$ number density profile
fits the observed slope of the surface brightness (see
below). We also fix the dust grain properties to be a collection of
compact spherical silicate grains from \citet{draine84}, with size
distribution $N(a)\propto a^{-3.5}$ from $a_{min}$ to
$a_{max}=10a_{min}$. Because the disk is assumed to be gas-free, each
grain size has its own temperature: smaller grains are hotter than
larger at a given distance from the star. Therefore, for each value of
$a_{min}$, we need to adjust $R_{in}$ to obtain a maximum dust
temperature in the range 60--80\,K, i.e., a peak dust emission around
40--50\,$\mu$m (Fig. 4). Once this is set, we can adjust the total dust mass to
fit the SED. For values of $a_{min}$ ranging from 0.125 to 8\,$\mu$m
(with a factor of 2 increment), we perform this 2-parameter adjustment
using a grid of 9 regularly-sampled values of $R_{in}$ and $M_d$ and
compute the associated $\chi^2$ value. 
The narrowness of the
disk thermal emission peak requires $a_{max}\lesssim10\,\mu$m (hence
$a_{min}\lesssim1\,\mu$m). Larger grains deviate from the
Rayleigh regime in the mid-infrared and result in a much broader
emission peak than is observed. By allowing a smaller value of $a_{min}$
and larger values of $a_{max}/a_{min}$, we do not find
a lower limit to $a_{min}$ from the analysis of the SED alone, down to at
least 0.1\,$\mu$m. Provided $R_{in}$ and $M_d$ are adjusted
accordingly, we always find satisfying fits to the SED.


We then explore the disk surface brightness in order to better
constrain $a_{min}$. Indeed, assuming $a_{max}/a_{min}$=10,
the scattering properties of the dust
populations considered here vary strongly with $a_{min}$. The albedo
at 0.606\,$\mu$m decreases from 0.82 to 0.65 as $a_{min}$ increases
from 0.125\,$\mu$m to 1\,$\mu$m, because of the depletion of the
highly reflective small grains. At the same time, the scattering
asymmetry parameter, $g = \langle \cos \theta \rangle$, increases from 0.65 to 0.89 because large
grains scatter preferentially forward. As a consequence, the models
with small values of $a_{min}$ can be rejected based on our ACS image:
simulated images of the disk with $a_{min}=0.125\,\mu$m are
$\sim$4 mag arcsec$^{-2}$ too bright compared to our
observations. Furthermore, with such a dust population, the predicted
flux ratio between the front and back side is low enough,
independently of the inclination, that it should have been detected in
our data set. To fit both the surface brightness of the disk and the
apparent front/back asymmetry, our model indicates that
$a_{min}>0.5\,\mu$m, implying a narrow grain size distribution
and justifying $a$ $posteriori$ our assumption of $a_{max}/a_{min}$ = 10

Figure\,\ref{fig4} illustrates our best fitting
model, which has $a_{min}=1\,\mu$m, $a_{max}=10\,\mu$m,
$R_{in}=60$\,AU, $M_d=10^{-7}
M_\odot$ and $L_{IR}/L_{\star}$=2.2$\times$10$^{-3}$. Assuming a
disk inclination $i=65-70\degr$, a power law fit to the modeled surface brightness profiles
results in a $\sim$-3.5 index and the observed front/back intensity
ratio at 2\farcs3 along the disk semi-minor axis is $\gtrsim10$. Both
figures are consistent with the observations. Subsequent models were
calculated with $a_{min}=0.75$ and 1.25\,$\mu$m to better constrain
this parameter. We obtained marginally acceptable surface brightness
fits in both cases and we conclude that $0.75\,\mu\mathrm{m} \leq
a_{min} \leq 1.25\,\mu\mathrm{m}$ in the disk surrounding HD 15745.
This result is consistent with a grain blowout size of $\sim$1 $\mu$m
calculated for silicate grains.
Associated uncertainties on the disk inner radius and total disk mass
are on the order of 10\%.
Our quantitative estimates of $a_{min}$ and $a_{max}$ depend on
the assumptions we made, but the overall conclusion (order of
magnitudes for grain sizes, hence inner radius and total dust mass,
and the need for a narrow size distribution) are robust.


\section{Discussion}
HD 15745 is now the fifth main sequence F star with
a debris disk resolved in scattered light.  Three of these
have ring-like structure, whereas HD 15745 shares a 
wide-disk architecture with HD 15115 \citep[F2V;][]{kalas07}.   
The three ring-like architectures are found
around HD 139664 \citep[F5V;][]{kalas06}, 
HD 181327 \citep[F5/F6;][]{schneider06},
and HD 10647 \citep[F9V;][]{ krs07}.  
At the present time, the dust properties have been modeled
only for HD 181327, with a Henyey-Greenstein scattering asymmetry parameter
$g=0.3\pm0.03$ \citep{schneider06}.  For debris disks
around stars of other spectral types, AU Mic (M1V) has a scattering asymmetry
parameter similar to that of HD 15745 \citep{graham07},
whereas most other debris disk, such as Fomalhaut \citep[A3V;][]{kalas05}
and HD 107146 \citep[G5V;][]{ardila04} have a low scattering asymmetry,
comparable to that of HD 181327.  Our present model suggests
a concentration of dust grains in the range 1$-$10 $\mu$m, which is consistent with
a wavy grain size distribution \citep[departure from a simple
power law; e.g.][]{campo94, thebault07}.  Small
grains do not exist in sufficient quantity to counteract the highly asymmetric
scattering phase function of the larger grains, whereas in other
debris disk systems the collisional cascade may be robust to the
smallest sizes, producing relatively symmetric scattering.
Further refinements for the properties of dust around HD 15745 and
other debris disks requires resolved imaging observations at other
wavelengths and with polarization.  Ultimately the explanation for
why one debris disk has a particular grain size distribution may
depend on differences in disk dynamical properties and the
composition of grains (e.g., volatile vs. refractory elements).  

Among the F stars discussed above, HD 15115 and HD 181327 may belong
to the Beta Pic Moving Group, with age $\sim$12 Myr \citep{zuck01}.  HD 15745, on the
other hand, may be a member of the Castor moving group based on its
Galactic space motion.  \citet{moor06} calculate 
($U,V,W$) = ($-16.5\pm1.1,-10.8\pm1.3, -10.7\pm0.7$) km s$^{-1}$
for HD 15745, consistent with
($U,V,W$) = ($-10.7\pm3.5, -8.0\pm2.4, -9.7\pm0.7$) km s$^{-1}$ given for
the Castor Moving Group \citep[CMG;][]{byn98}.  Among the CMG members,
HD 38678 ($\zeta$ Lep), another
well-known debris disk star has
a Galactic space velocity ($U,V,W$) = ($-14.4\pm3.6, -11.1\pm3.0, -8.4\pm1.8$) km s$^{-1}$
that is closest to that of HD 15745.  \citet{byn98} estimate that the
CMG has an age of 200$\pm$100 Myr, although \citet{zuck04} estimate
a younger age for HD 15745 (``30?'' Myr) based on its location on an A-star
H$-$R diagram and the relatively large dust optical depth.  
\citet{moor06} similarly conclude that nearly all stars
with $L_{IR}/L_{\star}>5\times$10$^{-4}$ are younger
than 10$^8$ yr.  The infrared excess for HD 15745 is an order of magnitude greater
than that of CMG members Fomalhaut, Vega, and $\zeta$ Lep.  
From the five CMG candidates that are F-stars, two
have have been detected at far-infrared wavelengths.  HD 119124 (F7.7V) has
$L_{IR}/L_{\star} = 3-$6$\times10^{-5}$, comparable to that of 
Fomalhaut and Vega \citep{trilling07}.  HD 130819 (F3V) has
$L_{IR}/L_{\star}=3\times10^{-4}$ \citep{chen05}, which is still a 
factor of seven less than that of HD 15745.  

Recent work demonstrates a large  spread in dust optical depths
at a given age  \citep{rhee07} which may be a consequence of the spread of initial
disk masses \citep{wyatt07}.  Therefore the high
optical depth of HD 15745 cannot uniquely discriminate
between a $\sim$10 Myr or a $\sim$100 Myr age.  
Assuming steady-state evolution, the dust optical depth
of HD 15745 could arise from a 10 Myr-old debris disk
that started with $\sim$10 M$_\earth$ of solid material,
or a 100 Myr-old disk that started with $\sim$100 M$_\earth$ 
of solid material.  The ambiguity is augmented by allowing
stochastic spikes in dust production that may occur in
a few per cent of debris disks.  \citep[e.g.][]{beichman06}.
A more precise age for HD 15745 therefore requires further
input using stellar properties.  
However, we note that the large geometric vertical
optical depth of our model ($\tau = 5.8\times10^{-3}$ at 60 AU radius)
implies a fiducial collisional destruction
timescale $t_\mathrm{c} \sim 1 / \Omega \tau \sim 10^5$\,yr at
60\,AU.  This is significantly shorter than the age estimates for HD 15745, 
implying that the circumstellar dust is collisionally
evolved.  Moreover,
the rapid collision timescale indicates that the radius
of the inner disk hole approximates the location of
the planetesimal belt that regenerates the observed dust,
instead of signifying the radius where grains inspiraling
from Poynting Robertson drag are trapped in resonances with planetary
bodies \citep{wyatt05}.

\section{Summary}

Using the $HST$ ACS coronagraph, we present the first images
of dust-scattered light from a debris disk surrounding the F2V star
HD 15745.  The fan-shaped morphology is consistent with
an azimuthally symmetric disk with asymmetric isophotes due
to an asymmetric dust scattering phase function combined with a 67$\degr$
inclination to the line of sight.  HD 15745 belongs in a category
of wide disks with a sensitivity-limited outer extent near 500 AU.  Our silicate grain
models for the SED and the disk surface brightness suggest an
inner disk radius of 60 AU, a somewhat narrow grain size distribution
between 1 and 10 $\mu$m, and a total disk mass of 10$^{-7}~M_{\odot}$.
Future observations should determine the optical and near-infrared
colors, as well as polarization, to further constrain grain properties.
High-resolution imaging of the inner part of the disk should reveal
the inner hole that in the model presented here has radius 60 AU, or 0.5$\arcsec$)
radius from the star projected on the sky.  Due to the asymmetric scattering
phase function, the inner boundary of the disk hole should appear
as a 1$\arcsec\times2\arcsec$ elliptical feature.  Our silicate-grain disk model predicts
a peak $V$-band and $H$-band surface brightnesses of 14.8 and 12.7 mag arcsec$^{-2}$,
respectively, or about 1 magnitude redder than the star itself.  Near-infrared data
would thereby test our present disk model that is dominated by
relatively large grains ($>$1 $\mu$m), where the albedo is not significantly diminished in the 
near-infrared, yet the near-infrared  scattering is more isotropic 
($g$=0.57 in $H$ vs. 0.89 in $V$) as the observing wavelength
approaches the grain size.

\acknowledgements
{\small
{\bf Acknowledgements:}  Support for GO-10896 was provided by 
NASA through a grant from STScI
under NASA contract NAS5-26555.  
We used the Gemini Observatory,
operated by AURA, under agreement with the NSF.
This work was supported by the NSF Science and Technology Center for Adaptive Optics, 
managed by  UCSC under agreement AST - 9876783, and by the Michelson Fellowship Program.
}

\clearpage

\begin{figure}
\epsscale{0.5}
\plotone{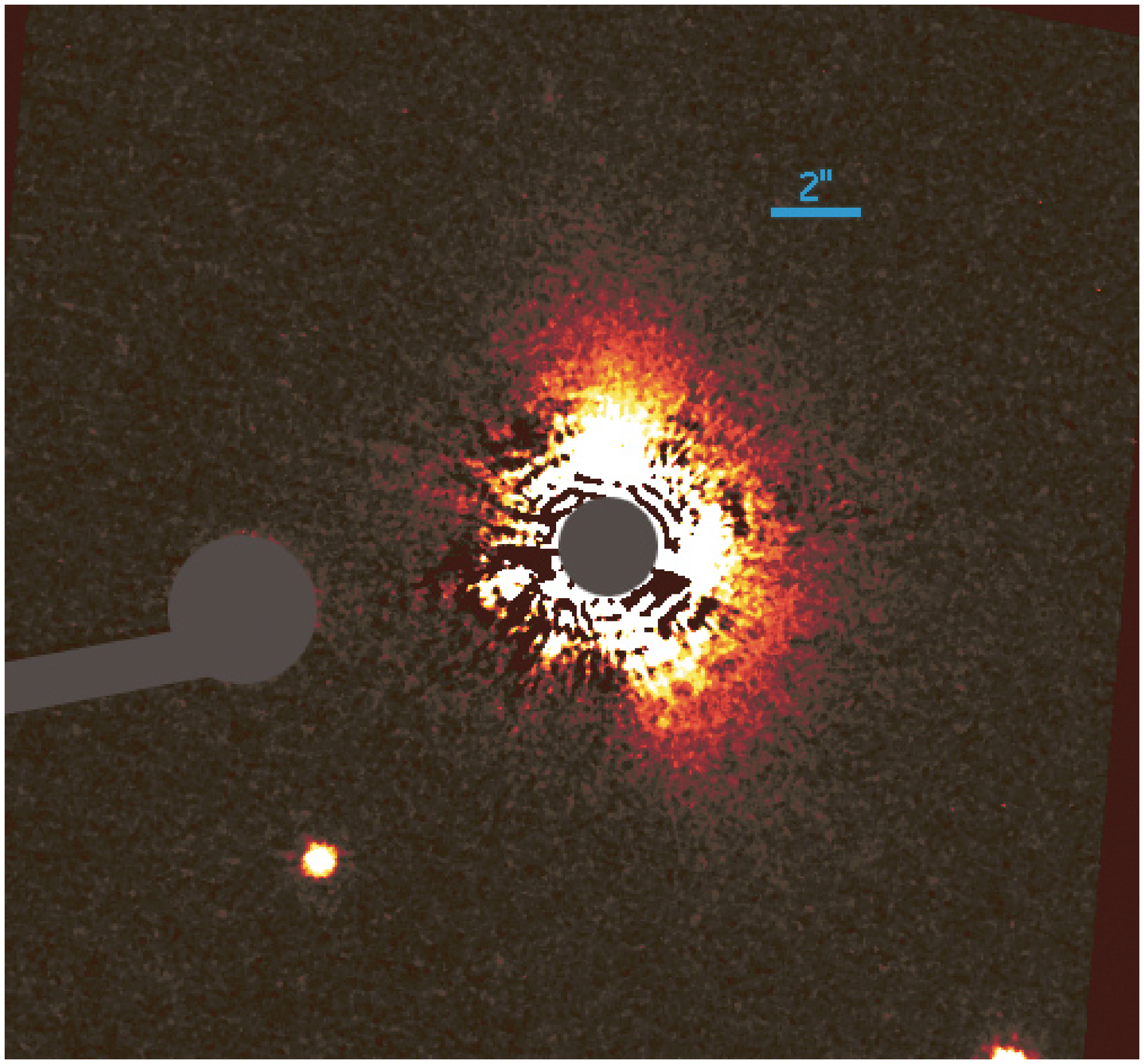}
\caption{
False-color, PSF-subtracted image of HD 15745.  North is up, east is left.
Observations were made using the F606W filter and the Advanced
Camera for Surveys High Resolution Channel.  
Gray fields cover the focal-plane masks in the ACS HRC.  The axis
of symmetry for the nebulosity is PA$\simeq$280$\degr$.
 \label{fig1}}
\end{figure}

\begin{figure}
\epsscale{0.5}
\plotone{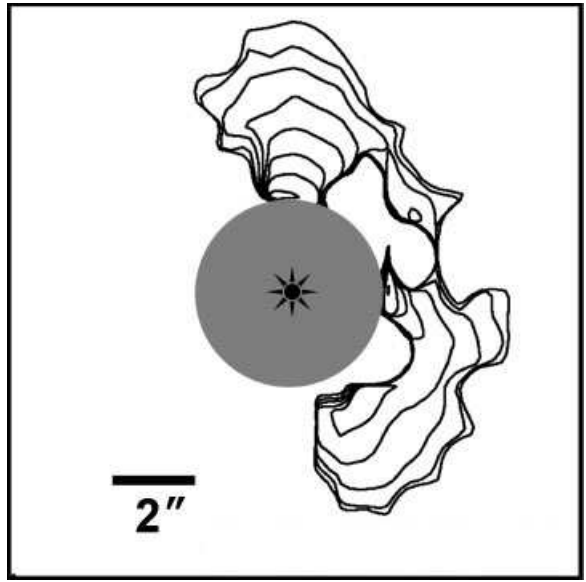}
\caption{
Surface brightness isocontours for the HD 15745 debris disk converted from F606W to
Johnson V-band (derived using STSDAS/SYNPHOT with a Kurucz model atmosphere and
the appropriate observatory parameters). 
North is up and east is left.  The outermost contour is 23.0 mag arcsec$^{-2}$,
with a contour interval of 0.5 mag arcsec$^{-2}$.
 \label{fig2}}
\end{figure}

\begin{figure}
\epsscale{0.75}
\plotone{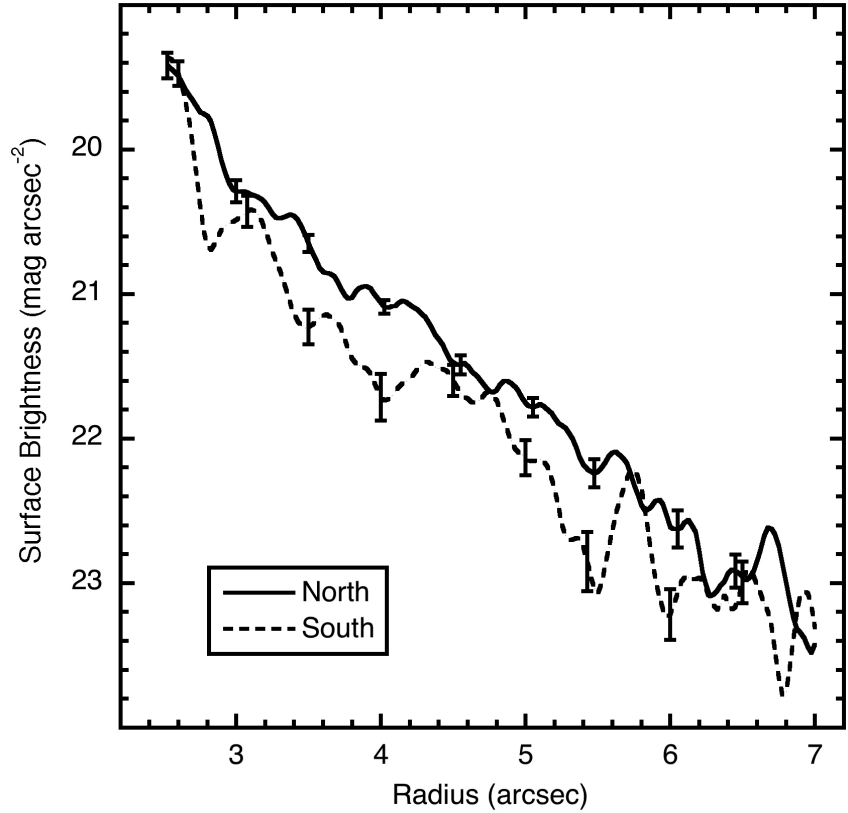}
\caption{
$V$-band radial surface brightness cuts between 2.5$\arcsec$ and 7.0$\arcsec$ radius
along the northern (PA=360$\degr$) and southern (PA=210$\degr$)
portions of the HD 15745 nebulosity.  We extracted photometry in boxes with sizes of 0.1$\arcsec$ (along
the radial direction) by 0.5$\arcsec$, and we plot the median value from the five subtractions of
the HD 15745 PSF.  Representative error bars show the standard deviation of these five values
for each box.  The noise due to PSF subtraction artifacts is greater for the south side of
the disk, as evidenced by the rippled form of the radial profile on scales of a few tenths of an arcsecond
and the larger error bars.  In some radial regions the south side of the disk may be fainter
than the north side by $\sim$0.4 mag arcsec$^{-2}$,
but these surface brightness asymmetries should be verified
in future data that have a higher signal-to-noise ratio.
\label{fig3} }
\end{figure}

\begin{figure}
\epsscale{0.75}
\plotone{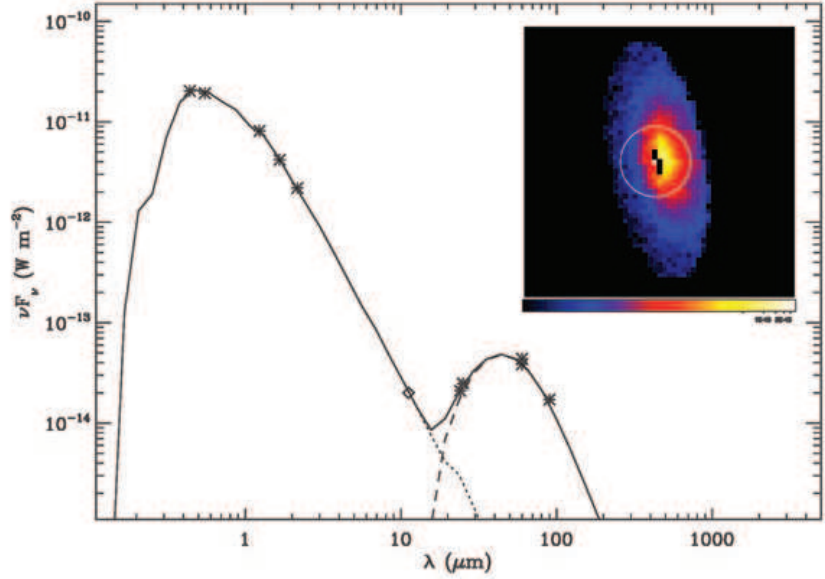}
\caption{
Observed SED of HD 15745 along with our best-fitting model (see text for model parameters). 
Our new N$^\prime$ ($\lambda_{c}$=11.2 $\mu$m, $\Delta\lambda$=2.4$\mu$m) Gemini measurement is indicated with a diamond (Gemini program ID:  GN-2006B-C-12).   
HD 15745 is unresolved with a
full-width at half-maximum of 0.3$\arcsec$ in the N-S direction,
and 0.5$\arcsec$ in the E-W direction.    Photometric calibration
was accomplished by observing HD 20893 in N$^\prime$ and 
assuming an extinction of 0.172 magnitude per air mass.  The 
24 $-$ 90 $\mu$m data represent the $IRAS$ and $ISO$ values given
by \citet{moor06}.
The dotted and dashed curves represent the stellar and disk emission, respectively. 
The inset is a log-stretch image of our best-fit model, observed with an inclination of 67$\degr$
and rotated so as to match the orientation of the disk in Fig. 1. The white circle represents the 2$\arcsec$
radius region that is hidden in Fig. 1 and 2.
\label{fig4} }
\end{figure}

\clearpage

\end{document}